\documentclass[journal,twoside,onecolumn,web]{ieeecolor}
\usepackage{generic}
\usepackage{cite}
\usepackage{amssymb,amsfonts}
\usepackage{algorithmic}
\usepackage{graphicx}
\def\BibTeX{{\rm B\kern-.05em{\sc i\kern-.025em b}\kern-.08em
    T\kern-.1667em\lower.7ex\hbox{E}\kern-.125emX}}
\usepackage{lmodern,enumerate,rotating,tensor}
\usepackage{amsmath,amsfonts,amssymb,mathtools,mathabx,blkarray,centernot}
\usepackage{tikz-cd}
\usetikzlibrary{cd}
\usetikzlibrary{shapes,shapes.geometric,arrows,fit,calc,positioning,automata}
\usepackage[normalem]{ulem}
\usetikzlibrary {circuits.logic.IEC}

\tikzset{
node distance=3cm, 
every state/.style={thick, fill=gray!10}, 
initial text=$ $, 
}

\tikzset{elliptic state/.style={draw,ellipse,thick, fill=gray!10},
initial text=$ $,
}

\tikzset{rectangular state/.style={draw, rectangle, thick, fill=gray!10}
}

\tikzset{emptystate/.style={}
}

\usepackage{stmaryrd}
\SetSymbolFont{stmry}{bold}{U}{stmry}{m}{n}

\newcounter{mycou}

\usepackage{pict2e}

\makeatletter
\newcommand{\CCOp}{\mathord{\mathpalette\nicoud@YESNO{\nicoud@path{\fillpath}}}}
\newcommand{\nicoud@YESNO}[2]{%
  \begingroup
  \settoheight{\unitlength}{$#1X$}%
  \begin{picture}(0.7,1)
  \linethickness{\variable@rule{#1}}%
  \roundcap\roundjoin
  \nicoud@path{\strokepath}
  #2
  \Line(0.35,-0.35)(0.35,1.08)
  \Line(0.55,-0.35)(0.55,1.08)
  \end{picture}%
  \endgroup
}
\newcommand{\nicoud@path}[1]{%
  \moveto(0.5,0.9)
  \lineto(0.5,1)\lineto(0.6,1)\lineto(0.6,0.9)
  \closepath
  \moveto(0.3,0.9)
  \lineto(0.3,1)\lineto(0.4,1)\lineto(0.4,0.9)
  \closepath
  \moveto(0.5,-0.27)
  \lineto(0.5,-0.17)\lineto(0.6,-0.17)\lineto(0.6,-0.27)
  \closepath
  \moveto(0.3,-0.27)
  \lineto(0.3,-0.17)\lineto(0.4,-0.17)\lineto(0.4,-0.27)
  \closepath
  #1
}
\newcommand{\variable@rule}[1]{%
  \fontdimen8  
  \ifx#1\displaystyle\textfont3\else
    \ifx#1\textstyle\textfont3\else
      \ifx#1\scriptstyle\scriptfont3\else
        \scriptscriptfont3\relax
  \fi\fi\fi
}
\makeatletter

\allowdisplaybreaks

\newcommand{\ep}{\epsilon}

\newcommand{\PSPACE}{\mathsf{PSPACE}}

\newcommand{\PTIME}{\mathsf{P}}

\usepackage{times,amsmath,amssymb,graphicx,tikz,algorithm,algorithmic,url,hhline,booktabs,soul}
\usetikzlibrary{automata,arrows,positioning}
\usetikzlibrary{petri}

\makeatletter
\newcommand{\subalign}[1]{%
  \vcenter{%
    \Let@ \restore@math@cr \default@tag
    \baselineskip\fontdimen10 \scriptfont\tw@
    \advance\baselineskip\fontdimen12 \scriptfont\tw@
    \lineskip\thr@@\fontdimen8 \scriptfont\thr@@
    \lineskiplimit\lineskip
    \ialign{\hfil$\m@th\scriptstyle##$&$\m@th\scriptstyle{}##$\crcr
      #1\crcr
    }%
  }
}
\makeatother

\makeatletter
\let\NAT@parse\undefined
\makeatother
\usepackage[colorlinks]{hyperref}

\definecolor{green}{rgb}{0.1,0.7,0.1}

\newcommand{\cyan}{\color{cyan}}

\newcommand{\ef}{{\cyan f}}

\usepackage{times,amssymb,amsfonts,mathtools,graphicx,tikz,algorithm,algorithmic,url,hhline,booktabs}
\usetikzlibrary{automata,arrows,positioning}

\usepackage{aliascnt}

\newaliascnt{lemma}{theorem}

\aliascntresetthe{lemma}

\newaliascnt{proposition}{theorem}

\aliascntresetthe{proposition}

\newaliascnt{corollary}{theorem}

\aliascntresetthe{corollary}

\newaliascnt{example}{theorem}

\aliascntresetthe{example}

\makeatletter
\newcommand{\mylabel}[2]{#2\def\@currentlabel{#2}\label{#1}}
\makeatother

\title{\LARGE A remark on diagnosability verification}
\author{Kuize Zhang, \IEEEmembership{Senior Member, IEEE}
  \thanks{Kuize Zhang is with Department of Mathematics and Statistics, Xi'an Jiaotong University, 710049 Xi'an, China (kuize.zhang@xjtu.edu.cn).}
}

\begin{document}

\maketitle

\begin{abstract}
    We point out three inaccuracies in paper [M.V. Moreira, T.C. Jesus, and J.C. Basilio. Polynomial time verification of decentralized diagnosability of discrete event systems. IEEE Transactions on Automatic Control, 56(7):1679--1684, July 2011]. First, the authors wrongly claimed that their algorithm for verifying (co-)diagnosability of labeled finite-state automata (LFSAs) did not depend on assumptions. We give an LFSA that is not deadlock-free or divergence-free such that their algorithm cannot correctly verify its diagnosability. Because diagnosability is a special case of co-diagnosability, their algorithm cannot correctly verify co-diagnosability either when LFSAs are not deadlock-free or divergence-free. 
  Second, they wrongly claimed that adding at each dead state an unobservable self-loop can help verifying diagnosability for an LFSA that is not deadlock-free or divergence-free, but this is wrong, because such a modification sometimes changes the diagnosability of an LFSA.
  Third, they wrongly claimed that their algorithm for verifying co-diagnosability ran in polynomial time. A polynomial-time algorithm unlikely exists, because the problem of verifying co-diagnosability of LFSAs is PSPACE-hard.
\end{abstract}



It was stated in \cite{Moreira2011Codiagnosability} that

\begin{equation*}
  \tag{A}\label{quoteA_comment_DiaVeri}
  \parbox{\dimexpr\linewidth-4em}{%
	\strut
	``We also remove, as in \cite{Qiu2006DecentralizedFD} and \cite{Wang2007DecentralizeDiagnosisDES},
	the assumptions of liveness of the language generated by the system
	and nonexistence of cycles of unobservable events.'',
	\strut
  }
\end{equation*}

and

\begin{equation*}
  \tag{B}\label{quoteB_comment_DiaVeri}
  \parbox{\dimexpr\linewidth-4em}{%
	\strut
	``\dots and assume, without loss of generality,
	that is live. If is not live, then there must be a deadlock state in the
	system, and so, can be made live by adding a self-loop labeled with
	an unobservable event \dots''.
	\strut
  }
\end{equation*}

We next show that the above statements are all inaccurate.

The verification method used in \cite{Moreira2011Codiagnosability} is an extension of the verifier 
proposed in \cite{Yoo2002DiagnosabiliyDESPTime} that depends on two assumptions of deadlock-freeness
(also called liveness) and divergence-freeness (that is, the LFSA does not generate cycles of unobservable events).
Note that it is a well known fact that the verifier method \cite{Yoo2002DiagnosabiliyDESPTime} depends on the two assumptions
(see Appendix of \cite{Zhang2023UnifiedFrame4DES}).
The co-diagnosability notions studied in 
\cite{Qiu2006DecentralizedFD} and \cite{Wang2007DecentralizeDiagnosisDES} are different from that
in \cite{Moreira2011Codiagnosability}, where the latter is
a direct extension of the one in \cite{Sampath1995DiagnosabilityDES}.

The verification method for co-diagnosability used
in \cite{Moreira2011Codiagnosability} does depend on the two fundamental assumptions of deadlock-freeness and
divergence-freeness, because the parallel composition is used. We next give an example to show
this fact.
The first verification algorithm for co-diagnosability that does not depend on
any assumption was given in \cite{Zhang2021UnifyingDetDiagPred}
based on a decentralized extension of the concurrent composition originally proposed in 
\cite{Zhang2019KDelayStrDetDES,Zhang2020DetPNFA} to verify the negations of strong versions of detectability,
where in \cite{Zhang2019KDelayStrDetDES,Zhang2020DetPNFA}, it is the first time that one can verify strong versions 
of detectability for arbitrary labeled finite-state automata. For the details on this series of
topics, we refer the reader to \cite{Zhang2023UnifiedFrame4DES}.

There are several issues in \cite{Moreira2011Codiagnosability}. First, the parallel composition $G_{N,1}||G_{N,2}||
G_{F}$ in Step~4 (bottom right corner of Page~1680) was not explicitly defined, so it was not explicitly shown what the 
verification structure is when there are at least two agents. That is, in \cite{Moreira2011Codiagnosability}, a verification algorithm for co-diagnosability was not given, only a
verification algorithm for diagnosability was given. Next, we give an example to show that the algorithm does 
depend on the two assumptions.

The necessary and sufficient shown in Step~5 (top left corner of Page~1681 of \cite{Moreira2011Codiagnosability}) is:
\begin{equation*}
  \tag{C}\label{quoteC_comment_DiaVeri}  
  \parbox{\dimexpr\linewidth-4em}{%
	\strut
	there is a (reachable) cycle in which there is a state with an $F$-component and after the state is an event.
	\strut
  }
\end{equation*}

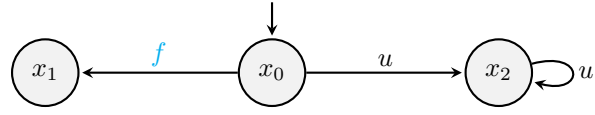
\begin{figure}[!htbp]
		\begin{center}
			\begin{tikzpicture}[
				>=stealth',shorten >=1pt,thick,auto,node distance=3.0 cm, scale = 1.0, transform shape,
	->,>=stealth,inner sep=2pt,
				every transition/.style={draw=red,fill=red,minimum width=1mm,minimum height=3.5mm},
				every place/.style={draw=blue,fill=blue!20,minimum size=7mm}]
				\tikzstyle{emptynode}=[inner sep=0,outer sep=0]
				\node[state, initial, initial where = above] (x0) {$x_0$};
				\node[state] (x1) [left of = x0] {$x_1$};
				\node[state] (x2) [right of = x0] {$x_2$};

				\path[->]
				(x0) edge node [above, sloped] {$\ef$} (x1)
				(x0) edge node [above, sloped] {$u$} (x2)
				(x2) edge [loop right] node {$u$} (x2)
				;
			\end{tikzpicture}
			\caption{LFSA $G$ (i.e., Fig.~23 of \cite{Zhang2023UnifiedFrame4DES}) , where $\ell(\ef)=\ell(u)=\ep$.}
			\label{fig1_High-OrderOpacity_ResponseLetter}
		\end{center}
	\end{figure}

Next, we use \autoref{fig1_High-OrderOpacity_ResponseLetter} to show why the verification method given in 
\cite{Moreira2011Codiagnosability} is generally incorrect when the two assumptions do not hold
even when there is a single agent. 
Obviously, $G$ is not deadlock-free (state $x_1$ is dead) or divergence-free ($x_2\xrightarrow[]{u} x_2$ is 
an unobservable cycle). We have shown that $G$ is $\{\ef\}$-diagnosable
in \cite[Appendix]{Zhang2023UnifiedFrame4DES}. One readily sees that $G$ satisfies Def.~1 of 
\cite{Moreira2011Codiagnosability} (corresponding to the case $m=1$ and $\Sigma_{o1}=\Sigma_o$). 
Following the notations of \cite{Moreira2011Codiagnosability},
$G_N$ and $G_F$ are shown in \autoref{fig2-1_High-OrderOpacity_ResponseLetter}
and \autoref{fig2-2_High-OrderOpacity_ResponseLetter}, respectively. The parallel composition
$G_N||G_F$ is shown in \autoref{fig3_High-OrderOpacity_ResponseLetter}, which satisfies 
\eqref{quoteC_comment_DiaVeri}. That is, the verification method of \cite{Moreira2011Codiagnosability}
wrongly showed that $G$ was not $\{\ef\}$-diagnosable. 
Therefore, statement \eqref{quoteA_comment_DiaVeri} is wrong.

	\begin{figure}[!htbp]
	  \centering
		  			\begin{tikzpicture}[
				>=stealth',shorten >=1pt,thick,auto,node distance=3.0 cm, scale = 1.0, transform shape,
	->,>=stealth,inner sep=2pt,
				every transition/.style={draw=red,fill=red,minimum width=1mm,minimum height=3.5mm},
				every place/.style={draw=blue,fill=blue!20,minimum size=7mm}]
				\tikzstyle{emptynode}=[inner sep=0,outer sep=0]
				\node[state, initial, initial where = above] (x0) {$x_0N$};
				\node[state] (x2) [right of = x0] {$x_2N$};

				\path[->]
				(x0) edge node [above, sloped] {$u$} (x2)
				(x2) edge [loop right] node {$u$} (x2)
				;
			\end{tikzpicture}
		  \caption{$G_N$.}
		\label{fig2-1_High-OrderOpacity_ResponseLetter}
	  \end{figure}
	  \begin{figure}[!htbp]
		\centering
			\begin{tikzpicture}[
				>=stealth',shorten >=1pt,thick,auto,node distance=3.0 cm, scale = 1.0, transform shape,
	->,>=stealth,inner sep=2pt,
				every transition/.style={draw=red,fill=red,minimum width=1mm,minimum height=3.5mm},
				every place/.style={draw=blue,fill=blue!20,minimum size=7mm}]
				\tikzstyle{emptynode}=[inner sep=0,outer sep=0]
				\node[state, initial, initial where = above] (x0) {$x_0N$};
				\node[state] (x1) [left of = x0] {$x_1F$};

				\path[->]
				(x0) edge node [above, sloped] {$\ef$} (x1)
				;
			\end{tikzpicture}
		  \caption{$G_F$.}
		  \label{fig2-2_High-OrderOpacity_ResponseLetter}
	\end{figure}

		\begin{figure}[!htbp]
		\centering
		  \begin{tikzpicture}[
				>=stealth',shorten >=1pt,thick,auto,node distance=4.5 cm, scale = 0.7, transform shape,
	->,>=stealth,inner sep=2pt,
				every transition/.style={draw=red,fill=red,minimum width=1mm,minimum height=3.5mm},
				every place/.style={draw=blue,fill=blue!20,minimum size=7mm}]
				\tikzstyle{emptynode}=[inner sep=0,outer sep=0]
				\node[elliptic state, initial, initial where = above] (x0Nx0N) {$(x_0N,x_0N)$};
				\node[elliptic state] (x0Nx1F) [left of = x0] {$(x_0N,x_1F)$};
				\node[elliptic state] (x2Nx0N) [right of = x0] {$(x_2N,x_0N)$};
				\node[elliptic state] (x2Nx1F) [below =1.5cm of x0] {$(x_2N,x_1F)$};

				\path[->]
				(x0Nx0N) edge node [above, sloped] {$\ef$} (x0Nx1F)
				(x0Nx0N) edge node [above, sloped] {$u$} (x2Nx0N)
				(x2Nx0N) edge [loop above] node {$u$} (x2Nx0N)
				(x0Nx1F) edge node {$u$} (x2Nx1F)
				(x2Nx0N) edge node {$\ef$} (x2Nx1F)
				(x2Nx1F) edge [loop below] node {$u$} (x2Nx1F)
				;
			\end{tikzpicture}

			\caption{$G_N||G_F$.}
			\label{fig3_High-OrderOpacity_ResponseLetter}
	\end{figure}

	  \begin{figure}[!htbp]
		\centering
			\begin{tikzpicture}[
				>=stealth',shorten >=1pt,thick,auto,node distance=3.0 cm, scale = 1.0, transform shape,
	->,>=stealth,inner sep=2pt,
				every transition/.style={draw=red,fill=red,minimum width=1mm,minimum height=3.5mm},
				every place/.style={draw=blue,fill=blue!20,minimum size=7mm}]
				\tikzstyle{emptynode}=[inner sep=0,outer sep=0]
				\node[state, initial, initial where = above] (x0) {$x_0N$};
				\node[state] (x1) [left of = x0] {$x_1F$};

				\path[->]
				(x0) edge node [above, sloped] {$\ef$} (x1)
				(x1) edge [loop left] node {$\ep$} (x1)
				;
			\end{tikzpicture}
		  \caption{$G_F'$.}
		  \label{fig4_High-OrderOpacity_ResponseLetter}
	\end{figure}

		\begin{figure}[!htbp]
		\centering
		  \begin{tikzpicture}[
				>=stealth',shorten >=1pt,thick,auto,node distance=4.5 cm, scale = 0.7, transform shape,
	->,>=stealth,inner sep=2pt,
				every transition/.style={draw=red,fill=red,minimum width=1mm,minimum height=3.5mm},
				every place/.style={draw=blue,fill=blue!20,minimum size=7mm}]
				\tikzstyle{emptynode}=[inner sep=0,outer sep=0]
				\node[elliptic state, initial, initial where = above] (x0Nx0N) {$(x_0N,x_0N)$};
				\node[elliptic state] (x0Nx1F) [left of = x0] {$(x_0N,x_1F)$};
				\node[elliptic state] (x2Nx0N) [right of = x0] {$(x_2N,x_0N)$};
				\node[elliptic state] (x2Nx1F) [below =1.5cm of x0] {$(x_2N,x_1F)$};

				\path[->]
				(x0Nx0N) edge node [above, sloped] {$\ef$} (x0Nx1F)
				(x0Nx0N) edge node [above, sloped] {$u$} (x2Nx0N)
				(x2Nx0N) edge [loop above] node {$u$} (x2Nx0N)
				(x0Nx1F) edge node {$u$} (x2Nx1F)
				(x0Nx1F) edge [loop above] node {$\ep$} (x0Nx1F)
				(x2Nx0N) edge node {$\ef$} (x2Nx1F)
				(x2Nx1F) edge [loop below] node {$u,\ep$} (x2Nx1F)
				;
			\end{tikzpicture}

			\caption{$G_N||G_F'$.}
			\label{fig5_High-OrderOpacity_ResponseLetter}
	\end{figure}

After adding a self-loop on state $x_1$ with event $u$, the new LFSA is not $\{\ef\}$-diagnosable
according to Def.~1 of \cite{Moreira2011Codiagnosability} (also corresponding to the case $m=1$ and $\Sigma_{o1}=\Sigma_o$).
This modification changes the diagnosability of $G$.
Therefore, statement \eqref{quoteB_comment_DiaVeri} is wrong.
Furthermore, if adding a self-loop on $x_1$ with a new event $\ep$ (the empty string),
this modified $G$ still satisfies Def.~1 of 
\cite{Moreira2011Codiagnosability} (corresponding to the case $m=1$ and $\Sigma_{o1}=\Sigma_o$), but the verification method of \cite{Moreira2011Codiagnosability} wrongly showed
that the modified $G$ was not $\{\ef\}$-diagnosable (the new $G_N||G_F'$ is shown in 
\autoref{fig5_High-OrderOpacity_ResponseLetter} and satisfies \eqref{quoteC_comment_DiaVeri}).

The co-diagnosability in \cite{Qiu2006DecentralizedFD} directly assumes that every terminating trace is diagnosable (see \cite[Def.~1]{Qiu2006DecentralizedFD}, ``or $st$ deadlocks''), so this definition is stronger than the
co-diagnosability 
in \cite{Moreira2011Codiagnosability}, and one can add at each dead state with an unobservable self-loop but does not change the co-diagnosability in \cite{Qiu2006DecentralizedFD} and \cite{Wang2007DecentralizeDiagnosisDES}. 
In other words, such an assumption set all traces whose diagnosability cannot be verified by using
the verifier to be diagnosable.

In addition, in \cite{Moreira2011Codiagnosability} the authors wrongly claimed that their exponential-time algorithm
as polynomial-time. A polynomial-time algorithm for verifying co-diagnosability unlikely exists because 
it is well-known that the co-diagnosability verification problem is PSPACE-hard in deadlock-free and
divergence-free LFSAs \cite{Cassez2012ComplexityCodiagnosability,Zhang2021UnifyingDetDiagPred}
and it is widely conjectured that $\PTIME\subsetneq \PSPACE$.


\bibliographystyle{unsrt}

\end{document}